\documentclass[preprint,showpacs,preprintnumbers,amsmath,amssymb]{revtex4}

\usepackage{graphicx}
\usepackage{dcolumn}
\usepackage{bm}

\begin{document}


\title{In-gap state and effect of light illumination in CuIr$_2$S$_4$ probed by photoemission spectroscopy}

\author{K. Takubo}
 \affiliation{Department of Physics and Department of Complexity Science and Engineering,
University of Tokyo, 5-1-5 Kashiwanoha, Chiba 277-8581, Japan}%
\author{T. Mizokawa}
 \affiliation{Department of Physics and Department of Complexity Science and Engineering,
University of Tokyo, 5-1-5 Kashiwanoha, Chiba 277-8581, Japan}%
\author{N. Matsumoto}
 \affiliation{
Department of Materials Science and Engineering, Muroran Institute of Technology, 27-1 Mizumoto-cho, Muroran,
Hokkaido, 050-8585 Japan}%
\author{S. Nagata}
 \affiliation{
Department of Materials Science and Engineering, Muroran Institute of Technology, 27-1 Mizumoto-cho, Muroran,
Hokkaido, 050-8585 Japan}%

\date{\today}

\begin{abstract}
We have studied disorder-induced in-gap states and 
effect of light illumination in the insulating phase 
of spinel-type CuIr$_2$S$_4$
using ultra-violet photoemission spectroscopy (UPS).
The Ir$^{3+}$/Ir$^{4+}$ charge-ordered gap appears 
below the metal-insulator transition temperature.
However, in the insulating phase, in-gap spectral features with 
$softgap$ are observed in UPS just below the Fermi level ($E_F$), 
corresponding to the variable range hopping transport observed 
in resistivity.
The spectral weight at $E_F$ is not increased by light illumination,
indicating that the Ir$^{4+}$-Ir$^{4+}$ dimer is very robust although 
the long-range octamer order would be destructed by the photo-excitation.
Present results suggest that the Ir$^{4+}$-Ir$^{4+}$ bipolaronic hopping and 
disorder effects are responsible for the conductivity of CuIr$_2$S$_4$.
\end{abstract}

\maketitle

\section{Introduction}
When some disorders are introduced in Mott insulators or charge-ordered insulators, 
disorder-induced electronic states are often observed within the band gaps and are 
responsible for the hopping transport in the insulating phase.
\cite{Nakatsuji,Husmann,Coey,Yoshida}
Although it is very important to understand the relationship 
between the transport behaviors and the disorder-induced in-gap states,
no unified picture on the relationship is obtained so far.
For example, in Ca$_{2-x}$Sr$_x$RuO$_4$, \cite{Nakatsuji}
NiS$_{2-x}$Se$_x$, \cite{Husmann} and R$_{1-x}$A$_x$MnO$_3$, \cite{Coey}
almost localized electronic states are formed within the band gap
and are responsible for the variable range hopping (VRH) transport.
On the other hand, in lightly-doped La$_{2-x}$Sr$_x$CuO$_4$\ ($x$=0.03), 
a sharp peak with clear band dispersion is observed within the Mott gap 
while the resistivity shows a VRH behavior. \cite{Yoshida}
The VRH transports and in-gap states suggest that the insulating
and metallic clusters coexist near the metal-insulator transition (MIT) in these materials. 
\cite{Pan,Sarma}
Such inhomogeneity plays crucial roles in their remarkable properties 
of colossal magnetoresistance and photo-induced MIT in manganites, 
\cite{Dagotto, Takubo} and stripe formation in cuprates. \cite{Kivelson}

CuIr$_2$S$_4$ with spinel structure is one of such systems and 
shows a MIT at $T_{MI} \sim$ 226K. \cite{first,jphys,optic2,high,optic,nmr,nmr2}
The anomalous hopping transport is also observed in the insulating phase
$\rho \propto $ exp[$-(T/T_0)^{1/2}$], \cite{resist,resist2}
or $\rho \propto$ $A$exp$(-E_a/k_BT)$ +$B$exp[$-(T/T_0)^{1/4}$]. \cite{Cao,hall}
The temperature variation of  $\rho \propto $exp[$-(T/T_0)^{1/2}$]
can be described by a Mott VRH conductivity in one dimensional case \cite{Mott} 
or by a Efros-Shklovskii VRH conductivity. \cite{Efros}
However, a rather complicated three dimensional charge ordering of Ir$^{3+}$ ($S$=0) and Ir$^{4+}$ ($S$=1/2) sites are indicated in the insulating phase 
of CuIr$_2$S$_4$. \cite{nature}
The cubic spinel structure of CuIr$_2$S$_4$ becomes tetragonally elongated
along the $c$-axis and bi-capped hexagonal ring octamers of 
Ir$^{3+}$ and Ir$^{4+}$ are formed below $T_{MI}$.
The orbital driven Peierls mechanism 
has been suggested, \cite{khomskii}
in which the Ir$^{4+}$ ions are dimerized along $xy$ chains of the $B$-sites.
Moreover, the resistivity of CuIr$_2$S$_4$ is
reduced by x-ray, \cite{xraymit1,xraymit2} visible light, \cite{PRL,vlight}
or electron beam irradiation \cite{tem,tem2} at low temperature ($\sim$ 100 K).
The symmetry of crystal is changed from triclinic to tetragonal
by the irradiation and the photo-induced state has a long lifetime.
It has been proposed that the photo-excitation breaks the Ir$^{3+}$/Ir$^{4+}$ 
charge ordering and induces the metallic conductivity.

The indication of MIT and the Ir$^{3+}$/Ir$^{4+}$ charge ordering
has been obtained on previous x-ray photoemission and absorption studies
of CuIr$_2$S$_4$. \cite{PRL,Noh,matsuno,xes,xas,cluster}
The Ir 4$f$ core-level spectrum of the insulating phase has two components 
with large energy difference, consistent with the charge ordering 
of Ir sites. \cite{PRL,Noh}
In contrast, the core-level spectrum has not been changed against 
laser irradiation, while the resistivity is reduced.

In this article, we report results of ultra-violet photoemission spectroscopy (UPS)
of CuIr$_2$S$_4$ single crystals combined with laser illumination. 
The UPS spectrum shows a clear MIT with band-gap opening $\sim$ 0.09 eV.
However, the UPS spectrum just below the Fermi level ($E_F$) at low temperature ($\sim$ 20 K) 
has a peculiar power-law dependence $\sim (E-E_F)^n$, $n \sim$ 1.3 - 1.7, 
that is associated with the anomalous conduction.
The systematic spectral change of UPS against laser irradiation has not been observed.
The Ir$^{4+}$-Ir$^{4+}$ dimers in the $xy$ plane of spinel are very robust 
against photo-excitation and play important roles in the exotic conduction.

\section{Experiments}
Single crystals of CuIr$_{2}$S$_{4}$ were grown by the bismuth solution method,
described previously in detail. \cite{crystal} 
UPS measurements were performed using SCIENTA SES-100 spectrometers
equipped with a He I source ($h\nu$ = 21.2 eV).
The total resolution was 30 meV and the base pressure of 
the spectrometer was $1\times10^{-7}$ Pa.
Five CuIr$_2$S$_4$ single crystals were studied for UPS measurements. 
The first and second samples named $\#R1$ and $\#R2$ are cleaved at 300 K 
\textit{in situ} and then measured at various temperatures.
The third, fourth, and fifth samples named $\#L3$, $\#L4$, and $\#L5$ 
are cleaved at 20 K \textit{in situ} and then measured at various temperatures.
All photoemission data were collected within 24 hours after the cleaving.
In order to study the effects of visible light excitation,
a Nd:YAG pulsed laser provided optical excitation to the samples 
at energies of 2.3 eV (532 nm) with a pulse frequency of 30 Hz
and a pulse width of about 10 ns. The beam was focused to a spot 
of 4 mm diameter.

\section{Results and discussions}

\subsection{Temperature dependence}

\begin{figure}
\begin{center}
\includegraphics[width=7cm,clip]{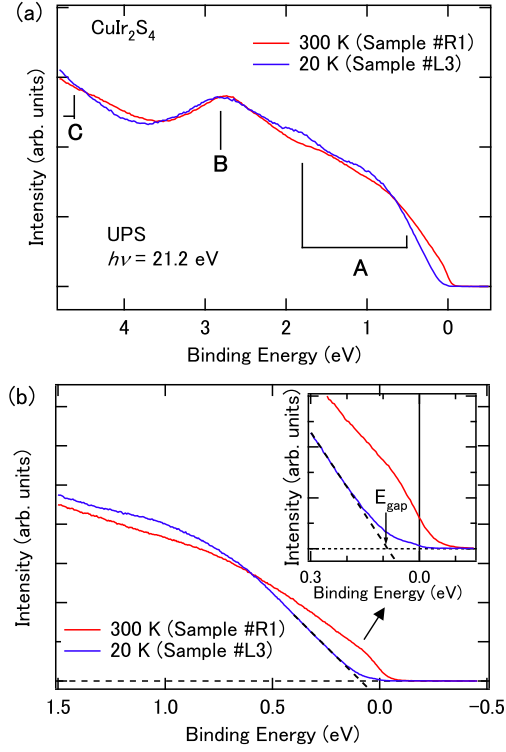}%
\end{center}
\caption{(Color online) UPS of CuIr$_2$S$_4$ taken at 300 K and 20 K immediately after
cleaving. (a) Wide-range spectra. (b) Near-$E_F$ spectra. Inset of (b) shows the expanded spectra near $E_F$.}
\label{v1}
\end{figure}

Figure \ref{v1} shows wide-range UPS of CuIr$_{2}$S$_{4}$ taken at 300 K for sample $\#R1$
and 20 K for sample $\#L3$ immediately after cleaving, respectively.
Compared with the previous studies, \cite{lda,lda2,xes,PRL,matsuno}
structures A, B, and C are assigned to the Ir 5$d$-S 3$p$ antibonding band, 
the Cu 3$d$ band, and Ir 5$d$-S 3$p$ bonding band, respectively.   
In the near-$E_F$ spectra, 
a spectral change across the MIT is observed.
The intensity at $E_{F}$ is substantial in
the spectrum at 300 K (sample $\#R1$, red line, metallic phase),
while the intensity at $E_F$ almost disappears in the spectrum at 20 K
(sample $\#L3$, blue line, insulating phase) [See Fig. \ref{v1} (b)].
The opening of the gap $E_{gap}$ $\sim$ 0.09 eV, obtained by extrapolating
the slope near the valance-band maximum to the base line,
agrees with the previous photoemission results \cite{PRL,Noh,optic} 
and is attributed to the Ir$^{3+}$/Ir$^{4+}$ charge ordering and 
Ir$^{4+}$-Ir$^{4+}$ dimerization along the $xy$ chains
with the tetragonal distortion.

\begin{figure}
\begin{center}
\includegraphics[width=7.5cm,clip]{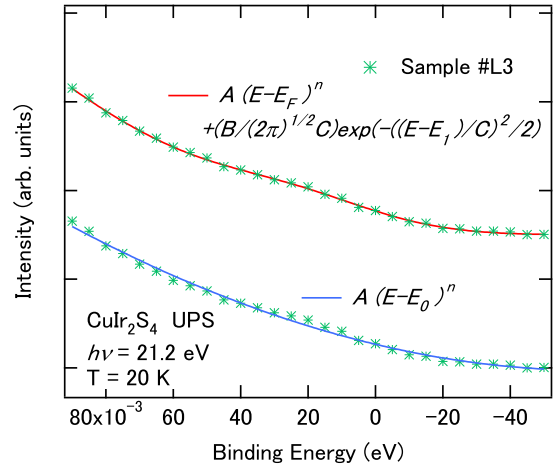}%
\end{center}
\caption{(Color online) UPS of CuIr$_2$S$_4$ for sample $\#L3$ taken at 20 K immediately after cleaving.
The solid curves indicate the fitted results with model functions
$A(E-E_F)^n$+$\frac{B}{\sqrt{2\pi}C}{\rm exp}[-(\frac{E-E_1}{\sqrt{2}C})^2]$ and $A(E-E_0)^n$, respectively.}
\label{f1}
\end{figure}

The spectrum of the insulating phase has a tail above the valence-band maximum
that reaches $E_F$ and forms a kind of $softgap$, interestingly.
This observation directly corresponds to the anomalous hopping transport
observed in the resistivity measurements
$\rho \propto $ exp[$-(T/T_0)^{1/2}$], \cite{resist,resist2}
or $\rho \propto$ $A$exp$(-E_a/k_BT)$ +$B$exp[$-(T/T_0)^{1/4}$]. \cite{Cao,hall}
We have fitted this in-gap spectral feature up to 0.09 eV ($\sim$ $E_{gap}$) 
to two types of model functions
$A(E-E_F)^n +\frac{B}{\sqrt{2\pi}C}{\rm exp}[-(\frac{E-E_1}{\sqrt{2}C})^2]$
and $A(E-E_0)^n$ (Fig. \ref{f1}).
The power-law function $A(E-E_F)^n$ is additionally convoluted with a
gaussian function taking into account the instrumental and thermal broadenings ($\sim$ 40 meV).
The gaussian peak $\frac{B}{\sqrt{2\pi}C}{\rm exp}[-(\frac{E-E_1}{\sqrt{2}C})^2]$ is assumed to be a contribution
from surface residual bonds near $E_F$ and it will be discussed later.
The integrated intensity is normalized to unity and
the fitted parameters are shown in Table \ref{t1}.  
The exponent of $n$ for sample $\#L3$ is estimated to 
be $\sim$ 1.5 and $\sim$ 1.7
by using the functions of $A(E-E_F)^n + \frac{B}{\sqrt{2\pi}C}{\rm exp}[-(\frac{E-E_1}{\sqrt{2}C})^2]$
and $A(E-E_0)^n$, respectively.

\begin{table}
\caption{
Best fit parameters of the model functions $A(E-E_F)^n + \frac{B}{\sqrt{2\pi}C}{\rm exp}[-(\frac{E-E_1}{\sqrt{2}C})^2]$
and $A(E-E_0)^n$ for the spectra of sample $\#L3$ of CuIr$_2$S$_4$.
The spectra are normalized using integrated intensity up to 0.09 eV.
}
\begin{center}
\begin{tabular}{ccccc}
\multicolumn{5}{c}{Function : $A(E-E_F)^n + \frac{B}{\sqrt{2\pi}C}{\rm exp}[-(\frac{E-E_1}{\sqrt{2}C})^2]$}\\
\hline\hline
Sample (20 K)& $n$ &  $B$ & $C$ (eV) & $E_1$ (eV) \\\hline
$\#L3$  & $1.5\pm0.2$  & $0.23\pm0.03$ & $0.02\pm0.01$ & $0.02\pm0.01$\\
\hline\hline
\\
\multicolumn{5}{c}{Function : $A(E-E_0)^n $}\\
\hline\hline
Sample (20 K) & $n$ & $B$ & $C$ (eV) & $E_1$ (eV) \\\hline
$\#L3$  & $1.7\pm0.2$  & - & - &-\\
\hline\hline
\end{tabular}
\label{t1}
\end{center}
\end{table}

\begin{figure*}
\begin{center}
\includegraphics[width=15cm,clip]{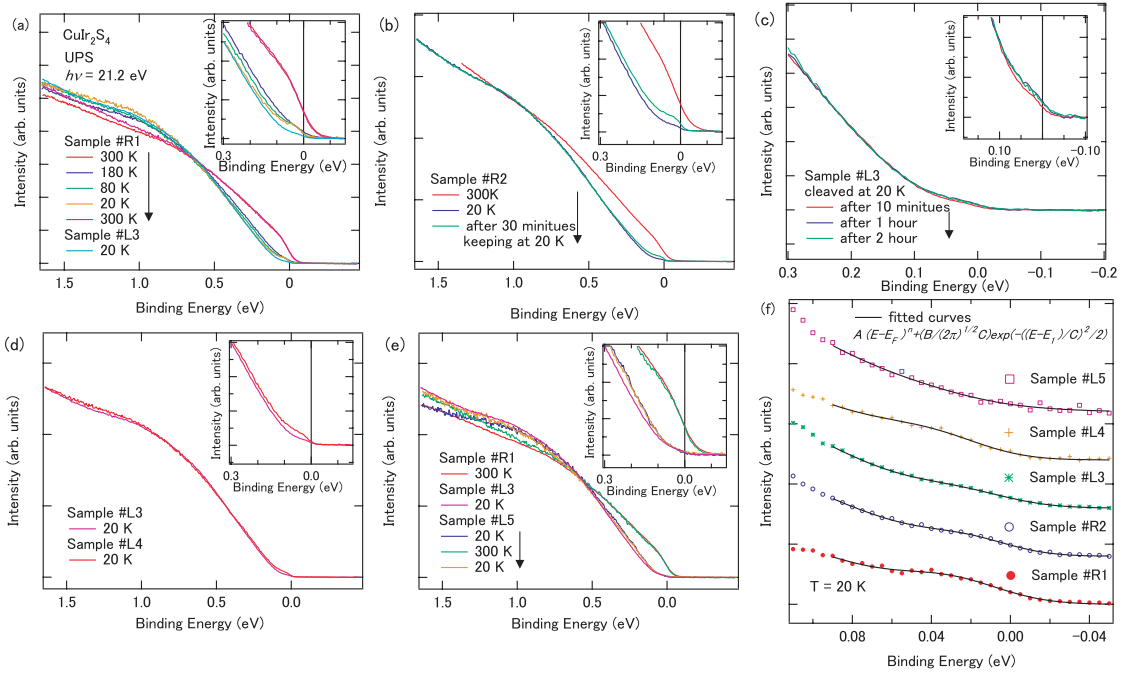}%
\end{center}
\caption{(Color online) UPS data of CuIr$_2$S$_4$ taken at various temperatures and for various samples.
(a)-(e) Insets are the expanded spectra near $E_F$. The arrows denote the order of the measurements.
(f) Near $E_F$ region of UPS data for CuIr$_2$S$_4$ taken at 20 K.
The solid curves indicate the fitted results with model functions
$A(E-E_F)^n +\frac{B}{\sqrt{2\pi}C}{\rm exp}[-(\frac{E-E_1}{\sqrt{2}C})^2]$.}
\label{v2}
\end{figure*}

Temperature dependence of the wide-range UPS was
carefully examined for various fractured surfaces.
Figure \ref{v2} (a) shows temperature dependence of UPS for sample $\#R1$,
which was cleaved at 300 K and then measured with decreasing temperature.
Although the spectra show the band-gap opening, a hump structure near $E_F$ 
remains even at the lowest temperature $\sim$ 20 K in contrast to 
the spectra for sample $\#L3$ measured immediately after cleaving at 20 K.
Sample $\#R2$ was also cleaved at 300 K and then measured with decreasing temperature.
However on the spectrum immediately taken after cooling down,
such hump is almost absent [Fig. \ref{v2} (b) blue line].  
Moreover after keeping 20 K for 30 minutes, the intense hump (green line) appeared also for this sample $\#R2$.
Even for sample $\#L3$, cleaved at low temperature, the intensity gradually increased with time although the intensity is rather small compared to the samples cleaved at 300K ($\#R1$, $\#R2$). Even at 20 K, the intensity of the hump depends on 
the cleavage. While, for sample $\#L4$ cleaved at 20 K, the hump was clearly observed, 
its intensity was very small for sample $\#L5$.
The hump structure is probably due to the surface state composed of 
unpaired Ir$^{4+}$ at the surface. 
Thus the hump structure can be affected by various surface conditions 
including treatments such as cleaving, change of temperature or irradiation 
from He I source.
We have tried to fit all the spectra taken at 20 K
with the function  $A(E-E_F)^n + \frac{B}{\sqrt{2\pi}C}{\rm exp}[-(\frac{E-E_1}{\sqrt{2}C})^2]$
as shown in Fig. \ref{v2} (f) and Table \ref{t2}.
The exponents of $n$ are universally estimated to be $\sim$ $1.3$ - $1.6$ by using this function at low temperature of CuIr$_2$S$_4$, independent of the gaussian contribution, namely the surface contribution. Therefore, we can safely conclude
that, while the hump near $E_F$ is the surface contributions, 
the in-gap state with the power law function $A(E-E_F)^n$ with $n$ $\sim$ $1.3$ - $1.6$ is derived from the bulk and is responsible for the VRH transport, $\rho \propto $ exp[$-(T/T_0)^{1/2}$]
(or $\propto$ $A'$exp$(-E_a/k_BT)$ +$B'$exp[$-(T/T_0)^{1/4}$]).

\begin{table}
\caption{
Best fit parameters of the model function  $A(E-E_F)^n + \frac{B}{\sqrt{2\pi}C}{\rm exp}[-(\frac{E-E_1}{\sqrt{2}C})^2]$
for the spectra of sample $\#R1$, $\#R2$, $\#L3$, $\#L4$, and $\#L5$ of CuIr$_2$S$_4$.
The spectra are normalized using integrated intensity up to 0.09 eV.
}
\begin{center}
\begin{tabular}{ccccc}
\multicolumn{5}{c}{Function :  $A(E-E_F)^n + \frac{B}{\sqrt{2\pi}C}{\rm exp}[-(\frac{E-E_1}{\sqrt{2}C})^2]$}\\
\hline\hline
Sample (20 K)& $n$ &  $B$ & $C$ (eV) & $E_1$ (eV) \\\hline
$\#R1$ & $1.4\pm0.2$  & $0.38\pm0.03$ & $0.03\pm0.01$ & $0.03\pm0.01$\\
$\#R2$ & $1.3\pm0.1$  & $0.22\pm0.03$ & $0.02\pm0.01$ & $0.02\pm0.01$\\
$\#L3$  & $1.5\pm0.2$  & $0.23\pm0.03$ & $0.02\pm0.01$ & $0.02\pm0.01$\\
$\#L4$   & $1.4\pm0.1$  & $0.30\pm0.03$ & $0.03\pm0.01$ & $0.04\pm0.01$\\
$\#L5$   & $1.6\pm0.2$  & $0.24\pm0.03$ & $0.04\pm0.01$ & $0.03\pm0.01$\\
\hline\hline
\end{tabular}
\label{t2}
\end{center}
\end{table}

In the situation forming $softgap$ due to electron-electron Coulomb
repulsion of Efros-Shklovskii type, the spectral 
function is usually characterized
by $A(E-E_F)^2$ dependence. \cite{Efros,Massey}
However CuIr$_2$S$_4$ has complicated three-dimensional 
charge order and has $hardgap$ $\sim$ 0.09 eV.
Probably, some disorder in the charge-ordered state 
is the origin of the in-gap spectral feature and 
the VRH transport.
Recently based on transport measurements of Ca$_{2-x}$Sr$_x$RuO$_4$,
Nakatsuji \textit{et al.} proposed that the hopping exponent of $\alpha$ $\sim$ 1/2
is universal feature of the disordered Mott system close to the metal-insulator transition
and reflects the emergence of disorder-induced localized electronic states
in the Mott-Hubbard gap. \cite{Nakatsuji}
The presence of some kind of disorder such as coexisting metallic clusters
in the insulating phase, which can be created near the first order MIT,
gives a strongly localized state and $\alpha$ $\sim$ 1/2.
A distinct in-gap state is also observed in UPS of
Ca$_{2-x}$Sr$_x$RuO$_4$. \cite{Sudayama}

\begin{figure}
\begin{center}
\includegraphics[width=6cm,clip]{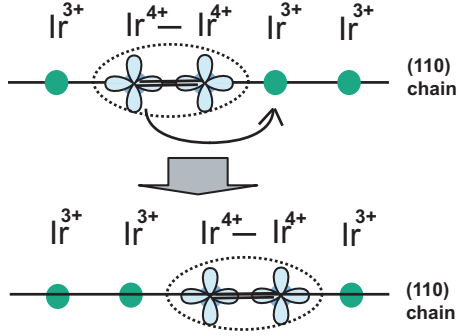}%
\end{center}
\caption{(Color online) Hopping of holes between an Ir$^{4+}$-Ir$^{4+}$ dimer and a neighboring Ir$^{3+}$ site along the (110) or (-110) chains.
The dimer moves without reducing the number of Ir$^{4+}$-Ir$^{4+}$ bonds. 
}
\label{chain}
\end{figure}

Furthermore, $n$ $\sim$ 1.3 - 1.7 for CuIr$_2$S$_4$ is rather small compared
to the expected value $\sim$ 2 of Efros-Shklovskii type.
The suppressed $n$ $\sim$ 1.5 of $A(E-E_F)^n$
is observed in a recent photoemission study on BaIrO$_3$, which has
a quasi-one-dimensional structure with Ir$_3$O$_{12}$ trimers
and also shows a charge-density wave transition. \cite{Maiti}
The exponent of $n$ = 1.5 for BaIrO$_3$ is attributed to the strong influence
of electron-magnon interaction.
The traveling dimer conduction in low temperature phase of CuIr$_2$S$_4$ is proposed
in Ref. \onlinecite{resist2}.
All Ir$^{4+}$ holes at low temperature fall in the $xy$ orbitals and 
form the dimers along the (110) [or (-110)] chains. \cite{khomskii}
When the Ir$^{4+}$-Ir$^{4+}$ dimer bonds are very strong and the system 
has good one dimensionality, single hole hopping from an Ir$^{4+}$-Ir$^{4+}$ dimer 
to a neighboring Ir$^{3+}$ site along the (110) chain may be suppressed.
Instead, paired hole hopping (bipolaronic hopping) of an Ir$^{4+}$-Ir$^{4+}$ pair
along the (110) chain is favored as shown in Fig. \ref{chain}.
The paired hole conduction may have quasi-one-dimensional and bipolaronic features
in the (110) or (-110) chains even in the three dimensional lattice.
Actually, the UPS line shapes of CuIr$_2$S$_4$
across the MIT resemble those of a bipolaronic material Ti$_4$O$_7$. \cite{Ti4O7}
The spectral weight in the insulating phase of Ti$_4$O$_7$
also obeys the power-law function $A(E-E_0)^n$ with $n$ $\sim$ 2.
Moreover, even in the metallic phase,
the both spectra of CuIr$_2$S$_4$ and Ti$_4$O$_7$ show weak Fermi edges
and broad peaks at $\sim$ 0.75 eV, 
around which most of spectral weights are distributed.
Such broad feature is commonly interpreted as the incoherent part
of the spectral function accompanying the quasi-particle excitations
around $E_F$. Probably, the fluctuation of the dimerization may 
survive at high temperature metallic phase.  

The temperature dependence of the UPS spectra across the MIT
is briefly summarized as follows.
The -Ir$^{3+}$-Ir$^{3+}$-Ir$^{4+}$-Ir$^{4+}$- charge ordering 
and Ir$^{4+}$-Ir$^{4+}$ dimerization
along the (110) or (-110) chains
cause the $hard gap$ opening of $\sim$ 0.09 eV, which 
also manifests in previous studies. \cite{PRL,xas,matsuno,Noh}
On the other hand, the peculiar in-gap state, which 
shows the power-law behavior with exponent $n$ $\sim$ 1.3 - 1.7,
gives the anomalous hopping conductivity at low temperature.

\subsection{Photo-excitation effects}

\begin{figure*}
\begin{center}
\includegraphics[width=15cm,clip]{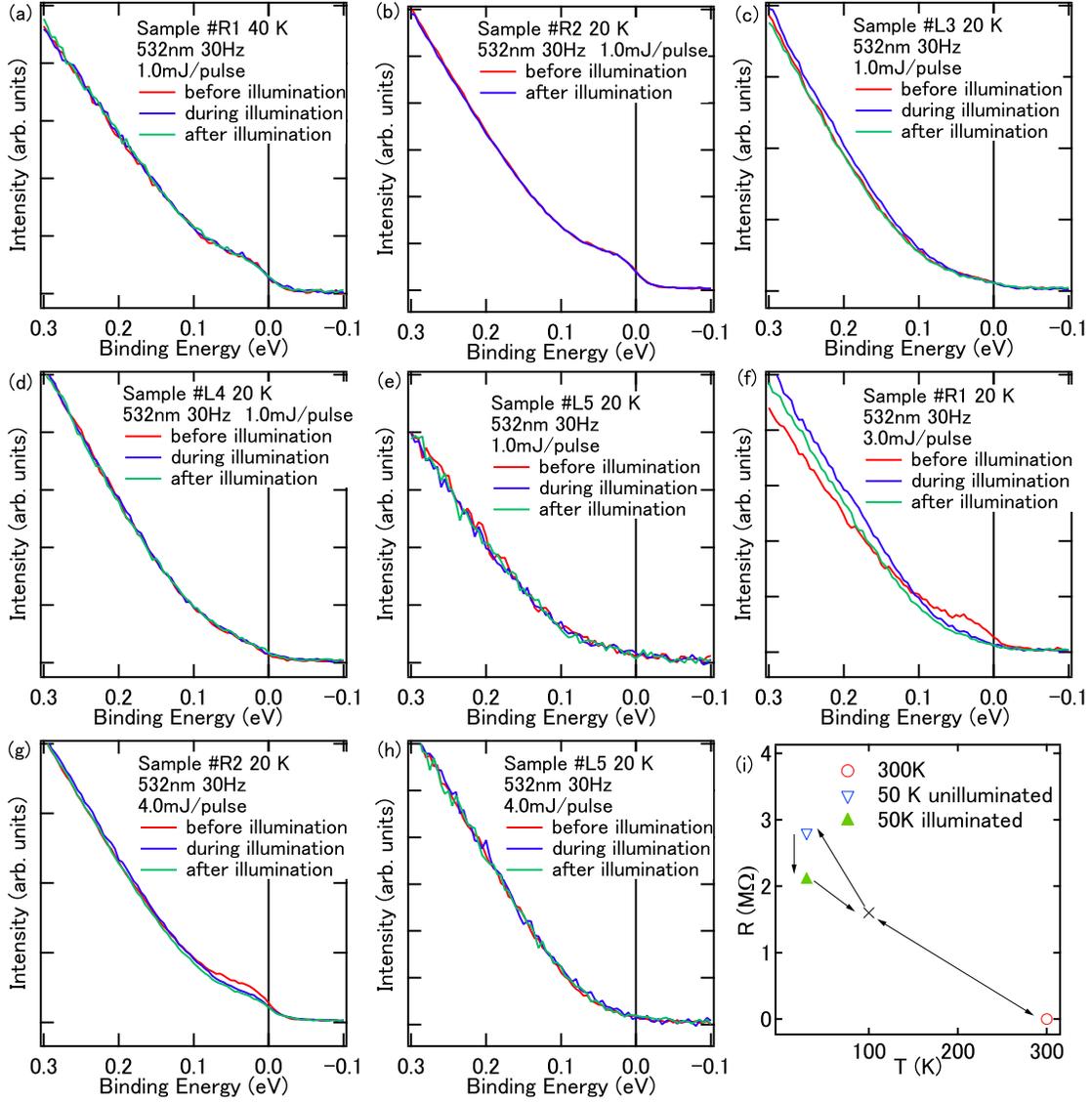}%
\end{center}
\caption{
(Color online) 
(a)-(h) UPS of CuIr$_2$S$_4$ before, during, and after visible light
irradiation from the Nd:YAG laser (532 nm) for various samples.
(i) The resistance taken under the same condition as the photoemission
measurements.
}
\label{l1}
\end{figure*}

We have also studied photo-excitation effects on the near-$E_F$ UPS spectra.
No spectral weight increase at $E_F$ was observed for any sample surfaces 
by visible light irradiation from the Nd:YAG laser up to 5 mJ/pulse
($8.5\times10^{16}$ cm$^2$photons/pulse) (See, Fig. \ref{l1}),
indicating that the Ir$^{3+}$/Ir$^{4+}$ charge ordered gap
in the insulating phase is very robust against photo-excitation
across the band gap. 
On the other hand, the present visible light irradiation also gives the reduction
of the resistance similar to those reported in the x-ray irradiation measurements
[Fig. \ref{l1} (i)]. The weak irradiation
up to 1 mJ/pulse ($1.7\times10^{16}$ cm$^2$photons/pulse)
gives no spectral change near $E_F$ [Fig. \ref{l1} (a)-(e)].
On the other hand, the hump of the in-gap states was decreased
by rather strong irradiations of $\sim 3$ - $5$ mJ/pulse [Fig. \ref{l1} (f)-(h)].
The strong irradiation may cause redistribution of charge at the
surface states.

\begin{figure}
\begin{center}
\includegraphics[width=8cm,clip]{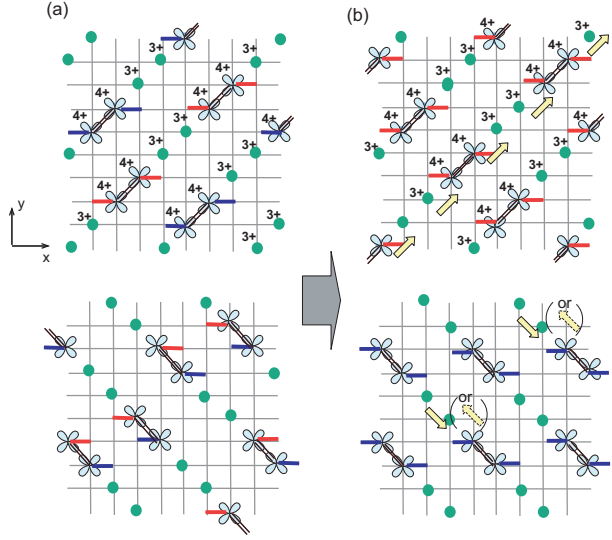}%
\end{center}
\caption{(Color online) (a) Charge ordering of Ir sites of CuIr$_2$S$_4$ sitting in one $xy$ plane.
The lower panel is another $xy$ plane a/4 below upper panel. 
The double solid lines denote the Ir$^{4+}$-Ir$^{4+}$ dimerized bonds.
The blue and red lines denote the octamer bonding toward Ir$^{4+}$ in the different planes.
(b) When the Ir$^{4+}$-Ir$^{4+}$ dimers shift by one unit,
the octamer ordering is destroyed and another order may appear.}
\label{phases}
\end{figure}

It has been suggested that the long-range charge ordering is
destroyed by the x-ray. \cite{xraymit1,xraymit2}
Probably, the visible light irradiation destroys
only the phase of -Ir$^{3+}$-Ir$^{3+}$-Ir$^{4+}$-Ir$^{4+}$-
chains in the $xy$ plane of spinel.
When the Ir$^{4+}$-Ir$^{4+}$ dimers in the $xy$ chains 
are shifted to the neighboring sites by the irradiation,
the octamer ordering is destroyed (See Fig. \ref{phases}).
This corresponds to the bipolaronic hopping as discussed in Fig. \ref{chain}
and can be regarded as a kind of bipolaronic solid-to-liquid
transition similar to Ti$_4$O$_7$. \cite{Ti4O7}
This picture is consistent with the recent diffraction study of
CuIr$_2$S$_4$ at low temperature, \cite{tem2}
indicating that the long range order (octamer order) is destroyed
but the dimers are preserved locally after x-ray or electron irradiation. 
Moreover similar charge ordering as shown in Fig. \ref{phases} (b) 
including -Rh$^{3+}$-Rh$^{3+}$-Rh$^{4+}$-Rh$^{4+}$- chains
is observed in the structural study of LiRh$_2$O$_4$,
which also has Rh sites at the $B$ site of spinel and shows 
a metal-insulator transition with some lattice distortion. \cite{LiRh2O4}

\section{Summary}
We have studied the electronic structure of CuIr$_2$S$_4$ 
single crystals using UPS. 
The UPS data shows the band-gap opening of $\sim$ 0.09 eV and 
supports the previous report of Ir$^{3+}$/Ir$^{4+}$ 
charge ordering in the insulating phase.
The observed in-gap state at low temperature is consistent with
the variable range hopping transport.  
The UPS measurements under laser irradiation indicate that the Ir$^{3+}$/Ir$^{4+}$ charge
disproportionation by the dimer formation is very robust against the photo-excitation
but the long-range charge ordering would be destroyed.

\section*{Acknowledgments}%
This work was supported by a Grant-In-Aid for Scientific Research
(Grants No.19340092) from the Ministy of Education, Culture, Sports, Science and Technology of Japan.
K. T. acknowledge support from the Japan Society for the Promotion of Science for Young Scientists.

\end{document}